\documentclass[review]{elsarticle}
\usepackage{hyperref}
\usepackage{amsmath}
\journal{Heliyon}
\begin{document}
\begin{frontmatter}
\title{Effects of experimental impairments on the security of continuous-variable quantum key distribution}
\author[label1]{Andres Ruiz-Chamorro\corref{corr}}
\ead{andres.ruiz@csic.es}
\cortext[corr]{Corresponding author}
\author[label1]{Daniel Cano}
%
\author[label1]{Aida Garcia-Callejo}
%
\author[label1]{Veronica Fernandez}
%
\address[label1]{Spanish National Research Council (CSIC), Institute of Physical and Information Technologies (ITEFI), Serrano 144, 28006 Madrid, Spain}
\begin{abstract}
Quantum Key Distribution (QKD) is a cutting-edge communication method that enables secure communication between two parties. Continuous-variable QKD (CV-QKD) is a promising approach to QKD that has several advantages over traditional discrete-variable systems. Despite its potential, CV-QKD systems are highly sensitive to optical and electronic component impairments, which can significantly reduce the secret key rate. In this research, we address this challenge by modeling a CV-QKD system to simulate the impact of individual impairments on the secret key rate. The results show that laser frequency drifts and small imperfections in electro-optical devices such as the beam splitter and the balanced detector have a negative impact on the secret key rate. This provides valuable insights into strategies for optimizing the performance of CV-QKD systems and overcome limitations caused by component impairments. By offering a method to analyze them, the study enables the establishment of quality standards for the components of CV-QKD systems, driving the development of advanced technologies for secure communication in the future.
\end{abstract}
\end{frontmatter}
%
\section{Introduction}
Quantum Key Distribution (QKD) is a secure communication method that enables two parties, Alice and Bob, to generate a secret key that is only known by them \cite{bennett_quantum_2014, pirandola_advances_2020, xu_secure_2020, renner_security_2008}. The key feature of QKD is its ability to detect the presence of any eavesdropper through the principles of quantum mechanics. The variables used to encode the quantum key are classified into two groups: continuous variables (CV), such as the quadratures of coherent states \cite{jouguet_experimental_2013, leverrier_finite-size_2010, diamanti_practical_2016, lodewyck_quantum_2007, grosshans_continuous_2002} and discrete variables (DV), such as the polarization states of single photons \cite{bennett_quantum_2014, ekert_quantum_1992, scarani_security_2009, lo_unconditional_1999, gisin_quantum_2002}. CV-QKD has several important advantages over DV-QKD, such as cost-effectiveness and ease of implementation \cite{grosshans_continuous_2002, zhang_integrated_2019}. In fact, CV-QKD does not require single-photon detectors, which makes it possible to use standard telecommunication devices, such as coherent receivers, instead. This presents an opportunity for the implementation of CV-QKD in current network infrastructures.

While the theoretical security of Quantum Key Distribution (QKD) is guaranteed by quantum principles, the quantum key rate of real experiments is highly sensitive to the impairments of optical and electronic components. Previous works have investigated the effects of these instrumental impairments or experimental imperfections \cite{silva_practical_2020,laudenbach_practical_2017, jouguet_analysis_2012, silva_role_2019}. However, a detailed analysis of individual impairments is still needed to fully understand the causes of decreased secret key rate in real CV-QKD systems, taking specific experimental imperfections into account. For example, imperfect basis choice and state preparation can reduce the security of the system by introducing inaccuracies in the estimation of channel properties \cite{liu_imperfect_2020,liu_imperfect_2017}. Additionally, receiver imperfections can impact the performance and security of CV-QKD systems, underscoring the importance of careful monitoring and compensation \cite{pereira_impact_2021}.

In this paper, we model a CV-QKD system to simulate the detrimental effects of individual impairments on the secret key. The model provides insights on optimization strategies to reduce excess noise and improve the secret key rate. The results are essential to establish the quality standards that each component must meet before being integrated into a CV-QKD network.
\section{Model system for CV-QKD}
\begin{figure}[t]
    \centering
    \includegraphics[width=\linewidth]{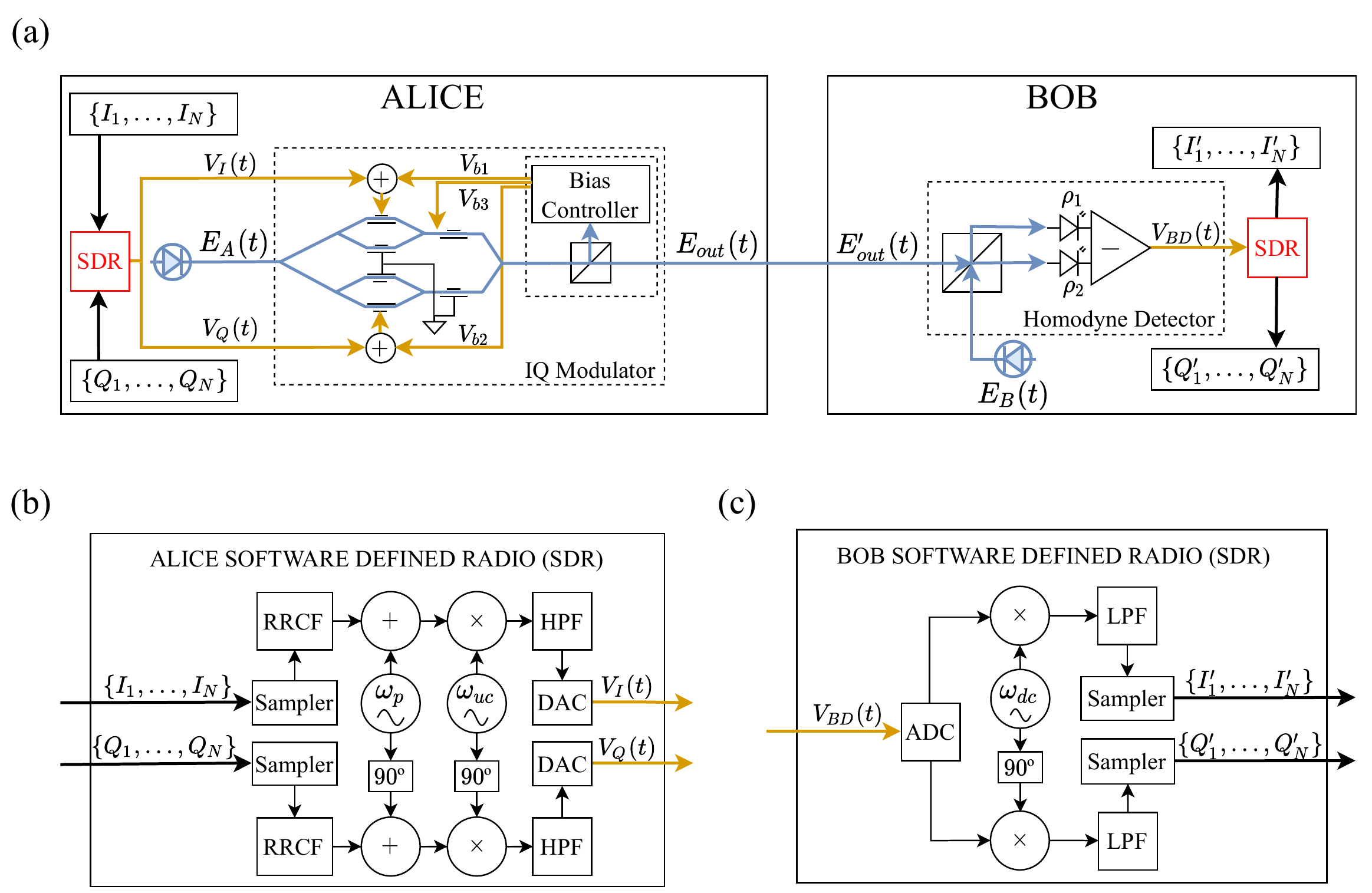}
    \caption{(a) Schematic of the CV-QKD system. Alice encodes the key in two sets of random values, $\{I_1,..., I_N\}$ and $\{Q_1,..., Q_N \}$. These are transformed into the signals $V_I(t)$ and $V_Q(t)$, which are used to modulate a laser field $E_A(t)$ in an IQ modulator. Bob carries out the low-complexity heterodyne detection using a local laser field $E_B(t)$, a beam splitter, and a balanced detector, whose output signal, $V_{BD}(t)$, is used to obtain $\{I_1',..., I_N'\}$ and $\{Q_1',..., Q_N' \}$. (b) Software Defined Radio in Alice. Two root-raised cosine filters (RRCF) smooth the symbols. Digital up-conversion is accomplished with a mixer and two high-pass filters (HPF). The analog signals are generated with two digital-to-analog converter (DAC). (c)  Software Define Radio in Bob. The analog signal is acquired in an analog-to-digital converter (ADC). Digital down-conversion is accomplished with a mixer and two low-pass filters (LPF).}
    \label{fig:setup}
\end{figure}

To study the effects of instrumental impairments on the secret key rate, we simulate a model system based on the most common and cost-effective implementations of CV-QKD \cite{brunner_low-complexity_2017}. In our system (see Figure \ref{fig:setup}), Alice sends a secret key by modulating the in-phase and quadrature components of weak coherent states using a single-mode laser and an IQ (In-Phase and Quadrature) modulator. The receiver (Bob) measures the signal using low-complexity heterodyne detection. This is the most cost-effective method implemented to date since it allows to obtain the in-phase and quadrature components by means of only one one detector in combination with a software post-processing of the signal \cite{brunner_low-complexity_2017}.

\begin{figure}
    \centering
    \includegraphics[width=\linewidth]{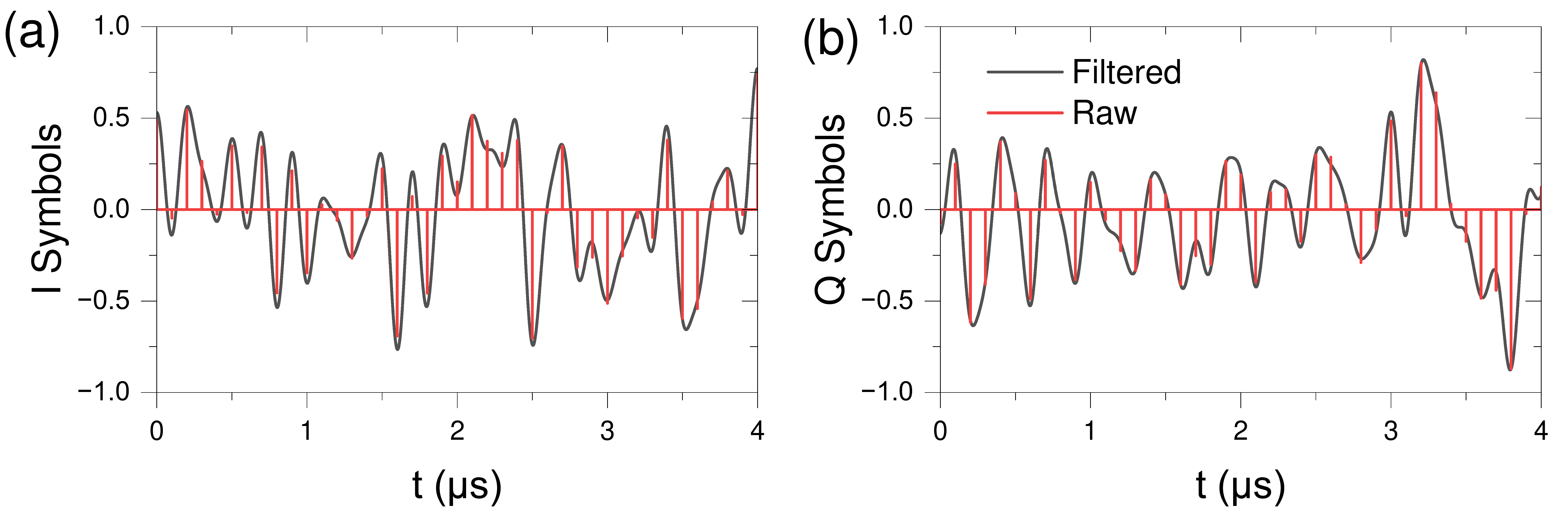}
    \caption{(a). In-Phase component of the symbols generated in Alice's Software Defined Radio before up-conversion. (b) Quadrature component of the same symbols. Note that all signals represented in this figure are obtained from numerical simulations and they are normalized for simplicity, as they could either be expressed in SI Units (Volts) or Shot Noise Units (SNU) \cite{laudenbach_continuous-variable_2018, zhang_improved_2020}.}
    \label{fig:sim_symbols}
\end{figure}

The basic operation of the system is described as follows. The protocol starts with the generation of a key, consisting of two sets of values, $\{I_1,I_2,...,I_N\}$ and $\{Q_1,Q_2,...,Q_N\}$, representing the in-phase and quadrature components of the coherent state modulation at a sampling rate $f_s$. To create hardware-implementable signals, each key string is filtered by a root-raised cosine filter (RRCF), producing two smooth signals as shown in Figure \ref{fig:sim_symbols}a and Figure \ref{fig:sim_symbols}b.

Since we use Gaussian modulation, the raw symbols are defined by $x\in\mathcal{N}(0,\sigma^2)$ at the sampling times and zero otherwise, being $\mathcal{N}(0,\sigma^2)$ a Gaussian distribution with zero mean and variance $\sigma^2$. The raw symbols are then filtered with the RRCF, generating the symbol signals that will be combined with a pilot tone, up-converted in frequency, and high-pass filtered before being sent to the IQ modulator.

A pilot tone of frequency $\omega_p$ is added to the symbol signals to provide a frequency reference and clock recovery method \cite{laudenbach_pilot-assisted_2019, wang_high-speed_2020}. Then, to avoid degradation of the low-frequency components, these signals are up-converted by mixing them with a signal of frequency $\omega_{uc}$, which is typically in the GHz domain. Two high-pass filters remove the redundant frequency components generated by the up-conversion mixing process. Finally, a digital-to-analog converter (DAC) generates the analog modulation signals, $V_I(t)$ and $V_Q(t)$, that are applied to the IQ modulator.  All the steps required to generate $V_I(t)$ and $V_Q(t)$ (and to demodulate $V_{BD}(t)$) are shown in Figure \ref{fig:setup}b (Figure \ref{fig:setup}c).

\begin{figure}
    \centering
    \includegraphics[width=\linewidth]{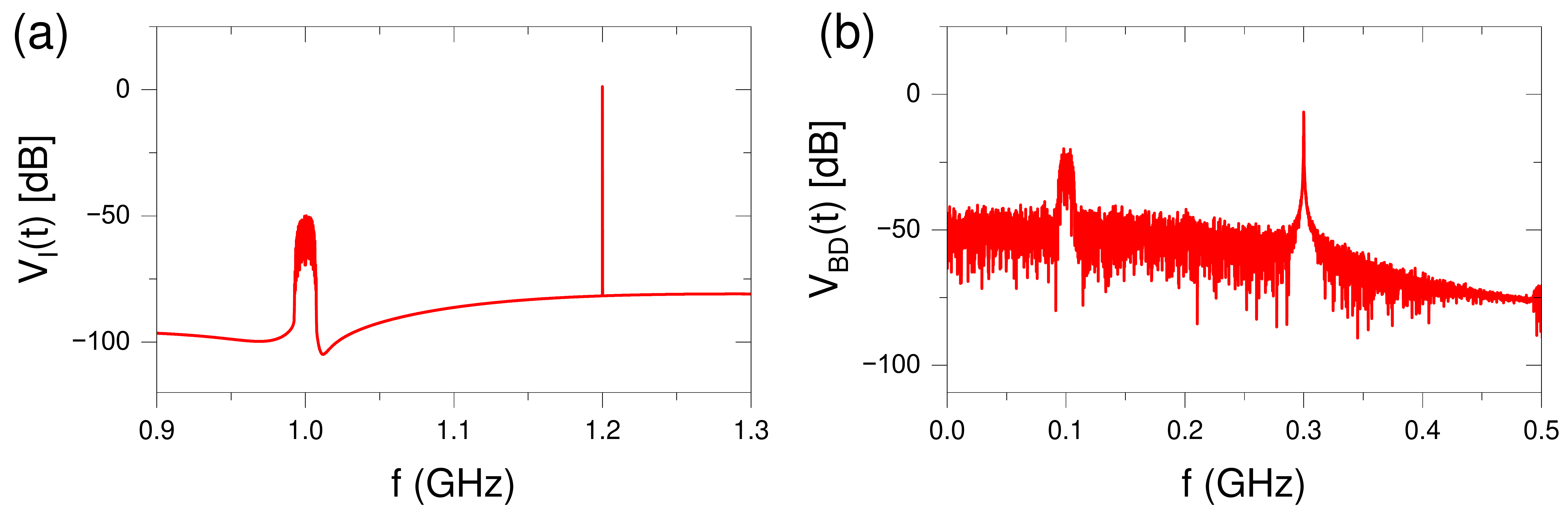}
    \caption{(a) The spectrum of the modulating signal $V_I(t)$.  (b) Spectrum of the balanced detector output signal $V_{BD}(t)$. Note that both signals are obtained from numerical simulations.}
    \label{fig:sim_sent_recv}
\end{figure}

We perform a numerical simulation of the aforementioned steps and present the results in Figure \ref{fig:sim_sent_recv}a. The left part of the spectrum displays the symbol band, which contains the information being transmitted. In addition to the symbol band, the right part of the spectrum is dominated by a single pilot tone, which is usually located far away from the symbol band to prevent it from interfering with the information being transmitted. It is important to note that the spectrum of the signal $V_Q(t)$ would be similar to that of $V_I(t)$.

In order to maximize the security of the quantum key transmission, the modulation amplitude of the symbols should be set to a low level to maintain the desired level of uncertainty in the quadrature measurements.

The signal is transmitted from Alice to Bob through an optical fiber, typically several kilometers long. Bob measures the in-phase and quadrature components of the quantum signal by means of a low-complexity heterodyne detection setup \cite{brunner_low-complexity_2017}. For this, the incoming laser field interferes with Bob's local oscillator in a beam splitter, and a balanced detector is employed to measure the difference between both outputs of the beam splitter. The output of the balanced detector reflects the combined effects of the modulating signal, the noise introduced by the channel, and the finite bandwidth of the detector itself, which in this case is assumed to be 400 MHz. It is important to note that the simulation of the output spectrum includes all of these factors, so that the true performance of the system can be accurately assessed. This procedure is numerically simulated step by step, as described in the next Section, to obtain the output signal of the balanced detector shown in Figure \ref{fig:sim_sent_recv}b.

\begin{figure}
    \centering
    \includegraphics[width=\linewidth]{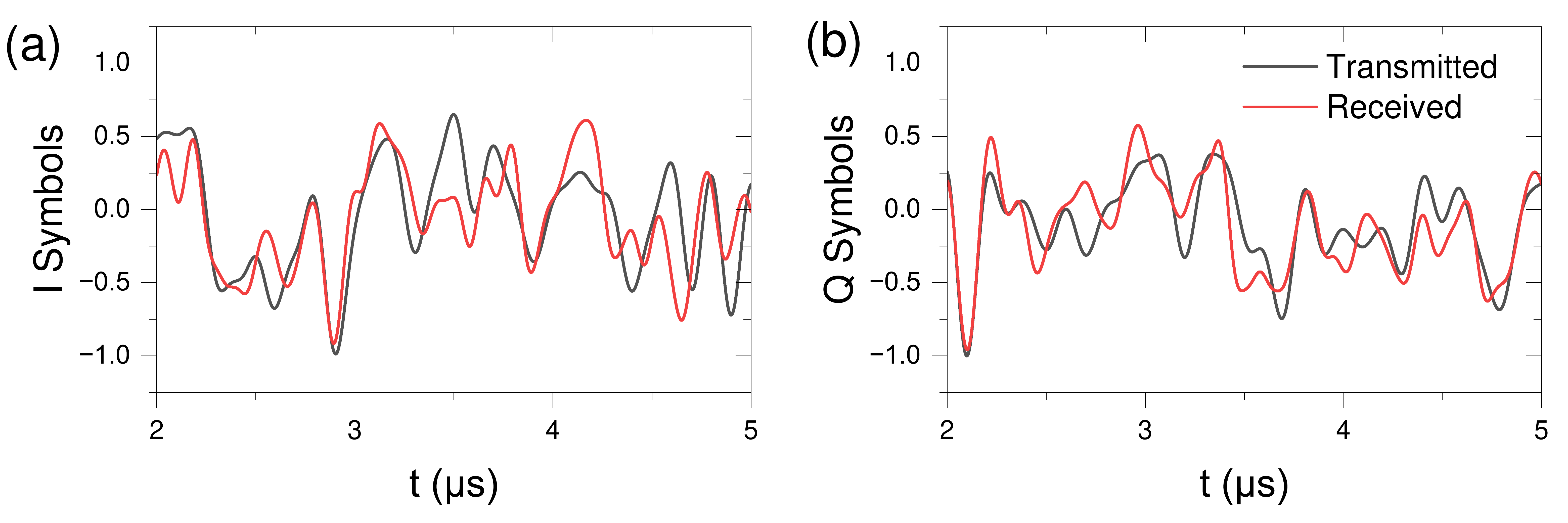}
    \caption{(a) In-Phase component of the symbol signal obtained in Bob's Software Defined Radio after low-complexity heterodyne detection (red) versus the symbol signal generated in Alice's Software Defined Radio (black). (b) Quadrature component of the same signals. Note that both signals are normalized such as in Figure \ref{fig:sim_symbols}.}
    \label{fig:sim_heterodyne}
\end{figure}

The signal is then down-converted using a frequency of $\omega_{dc}$ in combination with software post-processing that provides the in-phase and quadrature signals. Both in-phase and quadrature signals of Alice and Bob are shown in Figure \ref{fig:sim_heterodyne}a and Figure \ref{fig:sim_heterodyne}b. 
After recovering these signals, a parallel-to-serial encoder (P2S) is used to obtain two sets of random values, $\{I_1',I_2',...,I_N'\}$ and $\{Q_1',Q_2',...,Q_N' \}$, which should be correlated with those sent by Alice. In continuous-variable protocols, the security of the transmission can be estimated calculating the Secret Key Rate (SKR), which essentially depends on the difference between the mutual information shared between Alice and Bob, and the Holevo bound of the information shared between Eve and Bob, which represents the maximum amount of information that Eve could extract performing collective attacks on the channel \cite{leverrier_theoretical_2009,laudenbach_continuous-variable_2018,pirandola_advances_2020}.

To determine the secret key rate in the finite-size regime we reproduce the Parameter Estimation stage of the protocol using the algorithm of \cite{mountogiannakis_composably_2022}, where Alice and Bob randomly select a part of the transmitted data to study the correlation between the data strings that they have sent and received. This correlation allows them to compute the transmittance $T$ and excess noise $\xi$ of the channel. More details about Parameter Estimation and secret key rate calculation can be found in \cite{leverrier_theoretical_2009,leverrier_security_2013, renner_security_2006,laudenbach_continuous-variable_2018}

In the final steps of the protocol, Alice and Bob distill the key by means of privacy amplification and error correction algorithms \cite{devetak_distillation_2005}. The security of the transmission is characterized by its secret key rate, from which one can estimate an upper bound to the maximum information gained by a potential eavesdropper (Eve). If it exceeds the acceptable upper limit, Alice and Bob abort the protocol and discard the key. Otherwise, Alice and Bob perform error correction and privacy amplification techniques to finally distill a shared secure key.
\section{Instrumental impairments}
\subsection{Frequency drifts of the laser fields}
\begin{figure}
    \centering
    \includegraphics[width=0.75\linewidth]{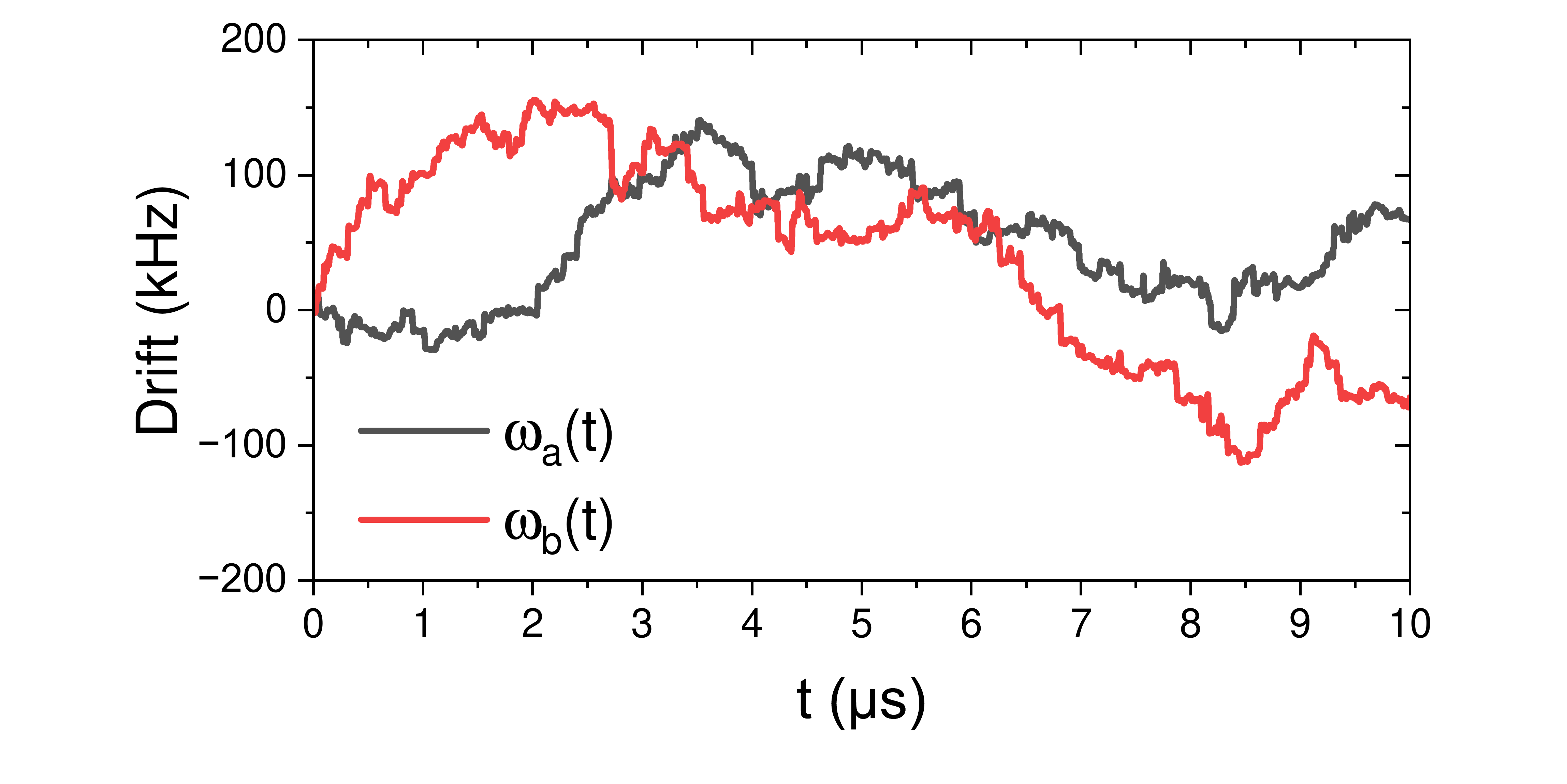}
    \caption{Simulation of a random drift in the frequencies of Alice and Bob's lasers, $\omega_A(t)$ and $\omega_B(t)$. We use a cubic interpolator to smooth the drift between two steps}
    \label{fig:sim_drift}
\end{figure}

Frequency drifts of the laser fields degrade the demodulation processes. Although small drifts can be partially corrected by off-line software post-processing \cite{soh_self-referenced_2015,wang_continuous-variable_2022}, their effect on the secret key rate  of the transmitted signal cannot be neglected. We simulate the frequency drifts by considering time-dependent frequencies, $\omega_A(t)$ and $\omega_B(t)$, for Alice and Bob's laser fields, which are respectively given by
\begin{equation}\label{eq:laser_drift}
    E_A(t) = |E_A| \exp(i\omega_A(t) t), \quad E_B(t) = |E_B| \exp(i\omega_B(t) t).
\end{equation}
In Eq. (\ref{eq:laser_drift}), the frequencies $\omega_A(t)$ and $\omega_B(t)$, which vary over time, are generated by random walks with step sizes based on a logistic distribution. To achieve a realistic simulation, the distribution was derived from experimental measurements of an external-cavity InP-based diode laser at 1550 nm. The simulation results are depicted in Figure \ref{fig:sim_drift}. 
\subsection{Stability of the bias controller in the IQ Modulator}
The IQ modulator is made up of a main Mach-Zehnder interferometer with two nested Mach-Zehnder interferometers, as shown in Figure \ref{fig:setup}a. Small deviations or noise in the input signals or the parameters of the three Mach-Zehnders can result in significant degradation of the final secret key rate. The output of the IQ modulator can be expressed as
\begin{equation}\label{eq:mzm_full}
    \small
    E_{\textrm{out}}(t) = \frac{E_{A}(t)}{2} \left[ \cos\left(\frac{\pi}{2} \frac{V_{b1} + V_I (t)}{V_{\pi1}}\right) + \exp\left(i \pi \frac{V_{b3}}{V_{\pi3}}\right) \cos\left(\frac{\pi}{2} \frac{V_{b2} + V_Q (t)}{V_{\pi2}}\right)\right],
\end{equation}
where $V_{b1}$, $V_{b2}$ and $V_{b3}$ are the bias voltages which set the operating points of each Mach-Zehnder interferometer, $V_{\pi1}$, $V_{\pi2}$ and $V_{\pi3}$ are their half-wave voltages; and $V_I(t)$ and $V_Q(t)$ are the analog modulation signals used to modulate each quadrature of the phase space. To implement the CV-QKD protocol, the bias voltages should be set to the quadrature operating point of the IQ modulator. This is achieved when $V_{b1}=-V_{\pi2}$, $V_{b2}=-V_{\pi2}$ and $V_{b3}=V_{\pi3}/2$. In this case, Eq. (\ref{eq:mzm_full}) is simplified to Eq. (\ref{eq:mzm_reduced}),
\begin{equation}\label{eq:mzm_reduced}
    E_{\textrm{out}}(t) = \frac{E_{A}(t)}{2} \left[\sin\left(m_1 V_I(t)\right) + i \sin\left(m_2 V_Q(t)\right) \right],
\end{equation}
where $m_i=\pi/2V_{\pi i}$ ($i=1,2$) are the modulation indexes that depend only on the half-wave voltage of the Mach-Zehnder modulators. In the ideal case, $m_1=m_2$ as $V_{\pi1}=V_{\pi2}$. It should be noted that in this analysis, we are specifically examining these particular electronic impairments. However, other studies \cite{liu_imperfect_2017, liu_imperfect_2020} have demonstrated how imperfect Gaussian state preparation can contribute to an increase in excess noise.
\subsection{Noise in the transmission channel}
For our simulations, we consider Gaussian noise as it represents the most comprehensive scenario for a transmission channel. This type of noise is generated by a class of attacks known as Gaussian attacks, which have been established as the most general attacks against CV-QKD \cite{pirandola_advances_2020, mountogiannakis_composably_2022}. The impact on the transmitted electric field can be thus described by Eq. (\ref{eq:eout}),
\begin{equation}\label{eq:eout}
    E_{\textrm{out}}'(t) = \sqrt{10^{-\alpha L/10}} E_{\textrm{out}}(t) +  x(t),
\end{equation}
where $L$ is the transmission channel length, $\alpha$ is the attenuation coefficient and $x(t)$ is a complex Gaussian noise signal. This signal is modeled as following a normal distribution with mean 0 and variance $\sigma^2$ at the symbol sampling times. The noise variance $\sigma^2$ is defined as the sum of the shot noise unit (1 SNU), the excess noise variance (the noise generated in the channel, both from attacks and imperfections, assumed to be 0.1 SNU), and the variance of the electronic noise (fluctuations in the electronic components that affect the measurement of the optical signals, assumed to be 0.01 SNU) \cite{mountogiannakis_composably_2022, laudenbach_continuous-variable_2018}.
\subsection{Incorrect ratio of the beam splitter}
The output fields of the beam splitter are given by \cite{leonhardt_quantum_2003},
\begin{equation}\label{eq:bsout}
      \left\{
    \begin{array}{l}
      E_1(t) = r_1 E_{\textrm{out}}'(t) + t_1 E_{B}(t) \\
      E_2(t) = t_2 E_{\textrm{out}}'(t) - r_2 E_{B}(t)
    \end{array},
  \right.
\end{equation}
where $r_1$ and $r_2$ are the reflection coefficients, and $t_1$ and $t_2$ are the transmission coefficients of the beam splitter. In the ideal case of a 50/50 beam splitter, we would have Eq. (\ref{eq:bsout}) with $t_1=t_2=r_1=r_2=\sqrt{1/2}$. Deviations from the ideal 50/50 beam splitter lead to reduction of the secret key rate. 
\subsection{Unbalanced detector gains}
The impact of the impairments of the balanced detector should also be taken into account. If the photo-detectors have different gains, the secret key rate also decreases. The output voltage is given by \cite{laudenbach_continuous-variable_2018}
\begin{equation}\label{eq:vdb}
    V(t) = \rho_1 P_1(t) - \rho_2 P_2(t).
\end{equation}
The parameters $\rho_1$ and $\rho_2$ ($P_1(t)$ and $P_2(t)$) in Eq. (\ref{eq:vdb}) are the sensitivities (powers) of the photo-detectors respectively. The signal measured at the balanced detector is shown in Figure \ref{fig:sim_sent_recv}b.
\section{Results and discussion}
In this section, we calculate the secret key rate including the effects of the instrumental impairments.  All simulations are carried out using the parameters of Table \ref{tab:parameters}. The values of the frequencies $\omega_A$, $\omega_B$, $\omega_p$, $\omega_{uc}$ and $f_s$ are standard values typically used in CV-QKD systems. The channel attenuation is the standard for single-mode telecommunication fibers ($\alpha=0.2$ dB/km). For $\eta$, $\nu$ and $\beta$ (a metric that represents the amount of information that is successfully extracted from the raw data), the values used are those from \cite{mountogiannakis_composably_2022} since they represent typical values for an experimental CV-QKD setup.
\begin{table}[t]
    \centering
    \begin{tabular}{c|c}
        Parameter & Value \\
        \hline
        Alice's laser frequency ($\omega_A$) & $2\pi \times 193.5000$ THz \\
        Bob's laser frequency ($\omega_B$) & $2\pi \times 193.5009$ THz \\ 
        Pilot tone ($\omega_p$) & $2\pi \times 200$ MHz \\
        Up-converting frequency ($\omega_{uc}$) & $2\pi \times 1$ GHz \\
        Symbol rate ($f_s$) & 10 MHz \\
        Channel attenuation & 0.2 dB/km \\
        Gaussian noise variance ($\sigma^2$) & 1.11 SNU \\
        Electronic noise variance ($\nu$) & 0.1 SNU \\
        Detection efficiency ($\eta$) & 0.8 \\
        Reconciliation efficiency ($\beta$) & 0.922 \\
    \end{tabular}
    \caption{Parameters used in the simulations.}
    \label{tab:parameters}
\end{table}

The theoretical model used for the computation of the secret key rate follows A. Leverrier's thesis \cite{leverrier_theoretical_2009}. The analysis here performed is based on the entanglement-based version of the protocol and calculates the secret key rate as a trade-off between the mutual information shared by Alice and Bob and the Holevo bound on the information that Eve shares with Bob. This is represented mathematically by
\begin{equation}\label{eq:skr}
    K = \beta I(A;B) - S(E;B),
\end{equation}
where $0\leq \beta \leq 1$ is the reconciliation efficiency and $S(E; B)$ is the Holevo information between Eve and Bob (i.e., an upper bound on the information Eve has on Bob’s data-measurement’s random variable). $I(A; B)$, as in the classical setting, represents the mutual information between Alice and Bob. The main objective –and challenge– of the security analysis is to compute the highest bound on Eve’s information of Bob's data, so as to obtain the lowest bound –and therefore safest- estimation of the secret key rate, defined as in Eq. (\ref{eq:skr}).

The Holevo bound can be computed considering a Gaussian state that shares with Alice and Bob’s state of information the same first two moments, i.e., the same covariance matrix. The desired covariance matrix, therefore, is computed from the parties data correlation statistics. After a few symmetrization procedures \cite{leverrier_theoretical_2009}, it takes the form: 
\begin{equation}\label{eq:covmatrix}
    \Gamma = \left(\begin{array}{cc}
    (V_A +1)I_2 & t_{min}Z\sigma_Z\\
    t_{min}Z\sigma_Z & (t_{min}^2+\sigma_{max}^2)I_2
    \end{array}\right),
\end{equation}
where $t_{min}$ and $\sigma_{max}^2$ are the minimal and maximal values, respectively, of the maximum-likelihood estimators (MLE) from which the real values of transmission $T$ and excess noise $\xi$ are approximated (ie., the worst-case estimations of the channel noise parameters). The expressions for the respective MLE's are given, respectively, by
\begin{equation}\label{eq:txi}
    \hat{t} = \frac{\sum_{i=1}^m x_iy_i}{\sum_{i=1}^m x_i^2},  \qquad  \hat{\sigma^2}  = \frac{1}{m}\sum_{i=0}^m (y_i - \hat{t}x_i)^2,
\end{equation}
being $m$ the number of states used to perform parameter estimation. The bound on the Holevo information for the Gaussian state with the covariance matrix of Eq. (\ref{eq:covmatrix}) is therefore given by a function of its eigenvalues $\nu_1$ and $\nu_2$, shown in Eq. (\ref{eq:holevo}),
\begin{equation}\label{eq:holevo}
    S(E;B) = \sum_{k=1}^3g\left(\frac{\nu_i-1}{2}\right),
\end{equation}
where $g(x) = (x + 1)\log_2(x + 1) - x \log_2 (x)$ and $\nu_3$ is the eigenvalue  the eigenvalue of the covariance matrix of Alice and Bob’s mode. By including this mathematical analysis in our simulations, we are able to calculate the secret key rate for multiple transmissions. This is computed after sampling the signals and obtaining the symbols $x_i$ and $y_i$, which are subsequently used to compute the transmittance and excess noise of the transmission using Eq. (\ref{eq:txi}), ultimately leading to the estimation of the secret key rate.
\begin{figure}[t]
    \centering
    \includegraphics[width=\linewidth]{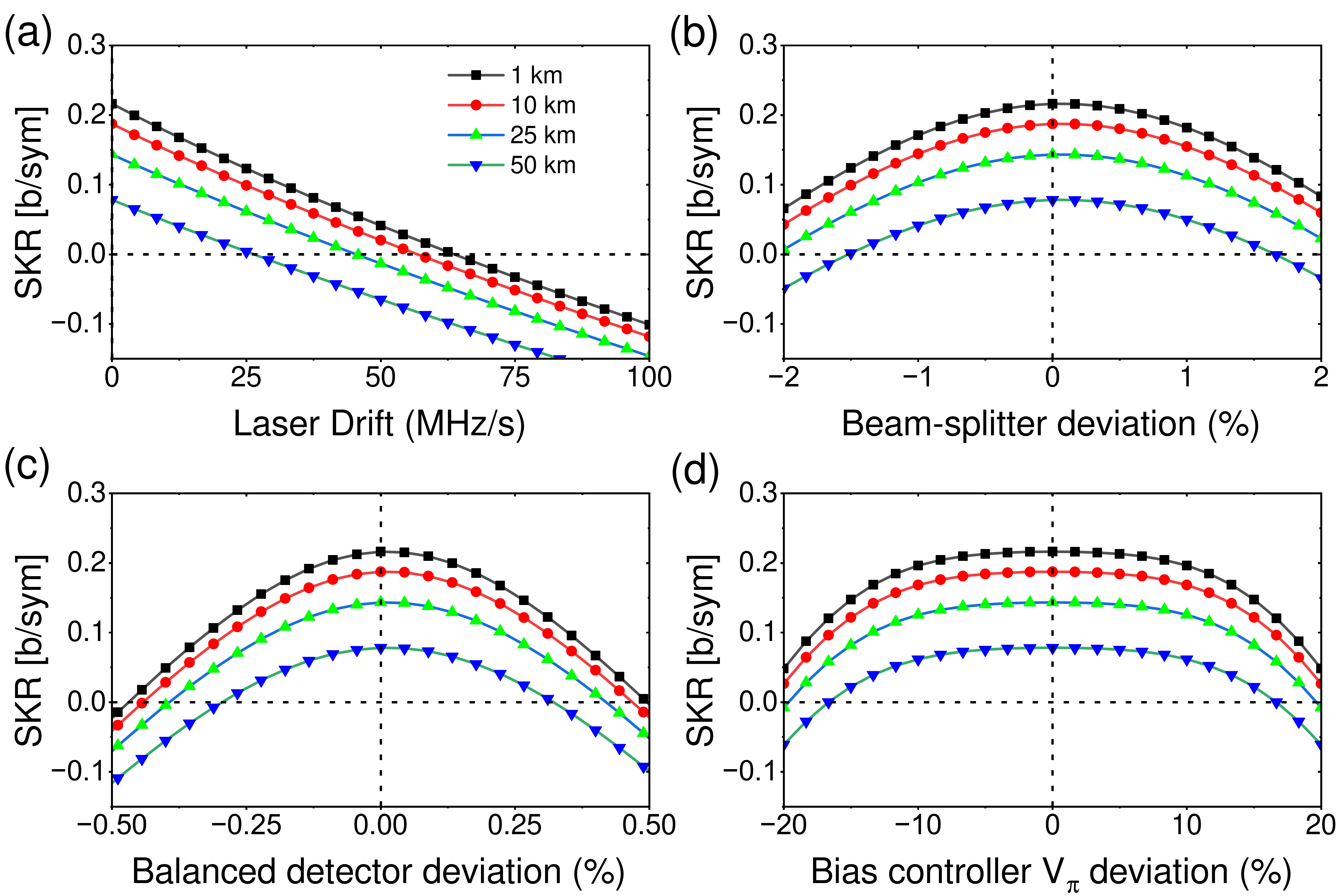}
    \caption{(a) Secret Key Rate versus deviations in reflectance and transmittance coefficients from its ideal value of $\sqrt{1/2}$, assuming $|r_1|^2+|t_1|^2=1$ and $|r_2|^2+|t_2|^2=1$. (b) Secret Key Rate as a function of the deviation of the photo-detector sensitivity $\rho_1$ with respect to $\rho_2$. (c) Secret Key Rate versus lasers maximum frequency drift per second. (d) Relation between secret key rate and the deviation of the bias controller from the ideal operating point, defined as $V_b=-V_\pi$, assuming $V_{b1}$ deviates from $-V_{\pi1}$. Very similar result is obtained for deviations in $V_{b2}$, $V_{b3}$.}
    \label{fig:results}
\end{figure}

Figure \ref{fig:results} shows the secret key rate as a function of instrumental impairments. One of the most noticeable effects is a dramatic decrease in the SKR due to the frequency drifts of both lasers, as shown in Figure \ref{fig:results}a. The reason is that these drifts introduce additional noise in the low-complexity heterodyne detection affecting the secret key rate. In fact, the SKR is below the security threshold for frequency drifts of just a few tens of MHz/s. The secret key rate is also very sensitive to small impairments in the beam splitter and the balanced detector, as shown in Figure \ref{fig:results}b and Figure \ref{fig:results}c. A deviation of only 0.1\% of the ideal $r$ or $t$ coefficients in the beam splitter, which is typical in standard beam splitters, can have a dramatic impact on the secret key rate. Interestingly, the SKR is hardly affected by small deviations of the bias voltage from its ideal operating point, as shown in Figure \ref{fig:results}d. This gives us some flexibility in the electronics used to stabilize the three interferometers of the IQ modulator by tuning its bias voltages.

In all figures, it can be observed that the secret key rate decreases with the distance between Alice and Bob, as expected. This is of particular relevance upon defining a range in which the characteristics of our experimental system must be found. As the distance increases, the range in which the SKR is positive becomes increasingly narrow, thus reducing the possibilities of transmitting a secure key. It is worth noting that standard commercial systems fall within these ranges. Lasers typically drift between 1 MHz/s and 50 MHz/s in frequency, beam-splitters typically deviate between 1\% and 5\% from the theoretical splitting ratio, balanced detectors usually have less than 1\% deviation between photodetector gains, and the error to determine the operating point in the IQ modulator bias controller is usually less than 5\%. This implies that for long distances, these features of the laser and beam-splitter will play a critical role in maintaining the secret key rate above a certain level to transmit a secure key.
\section{Conclusions}
In our study, we analyze the impact of different instrumental impairments on the Secret Key Rate (SKR) in a Continuous-Variable Quantum Key Distribution (CV-QKD) system. The results of the simulations show that laser frequency drifts can significantly lower the secret key rate and that even minor impairments in the beam splitter ratios or the gains of the photodiodes in the balanced detector can have a noticeable impact on key transmission. In contrast, small variations in the IQ (In-Phase and Quadrature) modulator's bias controller from its ideal operating point did not substantially affect the SKR.

Our model and simulations are a valuable tool in the initial calibration of a CV-QKD system, making it faster and easier, and preventing the decrease of the SKR due to most commonly-known impairments in experimental implementations. By determining and fixing the range of several physical parameters that enhance the transmission we can optimise the experimental performance of these systems, opening the possibility of long-distance or high-speed CV-QKD transmission.

Overall, this study presents how to enhance the SKR through a detailed theoretical model and simulation of the impairments present in experimental systems. It provides a new method to study each component of the setup independently and characterize, correct or mitigate the effect of several identified impairments. This contributes to the development and optimization of CV-QKD systems, which can play a crucial role in ensuring secure communication in future quantum networks.
\section*{Acknowledgments}
We would like to express our gratitude to Natalia Denisenko and Alfonso Blanco for their invaluable support and contributions to this research project. Their expertise and dedication greatly enhanced the quality of this publication.

This work had the support of Grant PID2020-118178RB-C22 funded by AEI/10.13039/501100011033, TED2021-130369B-C33 funded by MCIN/AEI/ 10.13039/501100011033 and by the European Union NextGenerationEU/PRTR, by the Community of Madrid (Spain) under the CYNAMON project (P2018/ TCS-4566), co-financed with European Social Fund and EU FEDER funds. We also acknowledge the support of CSIC’s Interdisciplinary Thematic Platform (PTI+) on Quantum Technologies (PTI-QTEP+). This study was supported by CSIC’s program for the Spanish Recovery,Transformation and Resilience Plan funded by the Recovery and Resilience Facility of the European Union, established by the Regulation(EU) 2020/2094; and MCIN with funding from European Union NextGenerationEU (PRTR-C17.I1).


\end{document}